\documentclass[aps,pra,reprint,groupedaddress,showpacs,longbibliography]{revtex4-1}

\usepackage{amssymb,amsmath,bm}
\usepackage{graphicx}
\usepackage{color}

\newcommand{\rr}{\mathbf{r}}
\newcommand{\DD}{\mathbf{d}}
\newcommand{\GG}{\mathbf{G}}
\newcommand{\ex}{\mathbf{e_x}}
\newcommand{\dd}{\mathrm{d}}

\makeatother

\begin{document}

\title{Transmission of near-resonant light through a dense slab of cold atoms}

\author{L. Corman$^\dagger$}
\author{J.L. Ville}
\author{R. Saint-Jalm}
\author{M. Aidelsburger$^\ddagger$}
\author{T. Bienaim\'e}
\author{S. Nascimb\`ene}
\author{J. Dalibard}
\author{J. Beugnon}

\email[]{beugnon@lkb.ens.fr}
\altaffiliation[$^\dagger$Present address: ]{Institute for Quantum Electronics, ETH Zurich, 8093 Zurich, Switzerland}
\altaffiliation[$^\ddagger$Present address: ]{Fakult\"at f\"ur Physik, Ludwig-Maximilians-Universit\"at M\"unchen, Schellingstr. 4, 80799 Munich, Germany}

\affiliation{Laboratoire Kastler Brossel, Coll\`ege de France, CNRS, ENS-PSL Research University, UPMC-Sorbonne Universit\'es, 11 place Marcelin-Berthelot, 75005 Paris, France}

\date{\today}
\begin{abstract}
\noindent 
The optical properties of randomly positioned, resonant scatterers is a fundamentally difficult problem to address across a wide range of densities and geometries. 
We investigate it experimentally  using a dense cloud of rubidium atoms probed with near-resonant light. The atoms are confined in a slab geometry with a sub-wavelength thickness.
We probe the optical response of the cloud as its density and hence the strength of the light-induced dipole-dipole interactions are increased. We also describe a theoretical study based on a coupled dipole simulation which is further complemented by a perturbative approach. This model reproduces qualitatively the experimental observation of a saturation of the optical depth, a broadening of the transition and a blue shift of the resonance.\\
\end{abstract}

\maketitle

\section{Introduction}

The interaction of light with matter is a fundamental problem which is relevant for simple systems, such as an atom strongly coupled to photons \cite{cohen1992atom, raimond2006exploring, pinkse2000trapping}, as well as for complex materials, whose optical properties provide information on their electronic structure and geometry \cite{ling2015renaissance}. This interaction can also be harnessed to create materials and devices with tailored properties, from quantum information systems such as memories \cite{gouraud2015demonstration}
and nanophotonic optical isolators \cite{sayrin2015nanophotonic} to solar cells combining highly absorptive materials with transparent electrodes \cite{britnell2013strong}.

The slab geometry is especially appropriate to study light-matter interaction \cite{damascelli2003angle, braun1996theory}. In the limit of a monolayer, two-dimensional (2D) materials exhibit fascinating optical properties. For simple direct band gap 2D semi-conductors, the single particle band structure implies that the transmission coefficient takes a universal value \cite{stauber2015universal, merthe2016transparency}. This was first measured for single layer graphene samples \cite{novoselov2004electric}, which have an optical transmission independent of the light frequency in the eV range, $|{\cal T}|^2=1-\pi \alpha$ where $\alpha$ is the fine structure constant \cite{nair2008fine,mak2008measurement}. 
The same value was recovered in InAs semiconductors \cite{fang2013quantum}. This universality does not hold for more complex 2D materials, for instance when the Coulomb interaction plays a more important role \cite{eda2013two}.

Atomic gases represent in many respects an ideal test bed for investigating light-matter interaction. First, they can be arranged in regular arrays \cite{bettles2016enhanced, shahmoon2017cooperative} or randomly placed \cite{Chomaz12} to tailor the optical properties of the system. Second, an atom always scatters light, in contrast with solid-state materials where the optical excitation can be absorbed and dissipated in a  non-radiative way. Even for thin and much more dilute samples than solid-state systems, strong attenuation of the transmission can be observed at resonance. Third, inhomogeneous Doppler broadening can be made negligible using ultracold atomic clouds. Finally, the geometry and the density of the gases can be varied over a broad range. 

In the dilute limit, such that the three-dimensional (3D) atomic density $\rho$ and the light wavenumber $k$ verify $\rho k^{-3}\ll1$, and for low optical depths, a photon entering the atomic medium does not recurrently interact with the same atom. Then, for a two-level atom, the transmission of a resonant probe beam propagating along the $z$ axis is given by the Beer-Lambert law: $|{\cal T}|^2= e^{-\sigma_0 \int \rho dz}$, where $\sigma_0=6\pi k^{-2}$ is the light cross section at the optical resonance \cite{Yefsah2011exploring}. At larger densities the transmission is strongly affected by the light-induced dipole-dipole coupling between neighboring atoms.

Modification of the atomic resonance lineshape or super- and sub-radiance in dilute (but usually optically dense) and cold atomic samples have been largely investigated experimentally \cite{bienaime2010observation, balik2013near, meir2014cooperative, kemp2014cooperatively, kwong2014cooperative, goban2015superradiance, araujo2016superradiance, roof2016observation, guerin2016subradiance,Bromley16}. Recently, experiments have been performed in the dense regime studying  nanometer-thick hot vapors \cite{Keaveney2012} and mesoscopic cold clouds \cite{pellegrino2014observation,jenkins2016optical, jennewein2016, jenkins2016collective}. Interestingly, it has been found that the mean-field Lorentz-Lorenz shift is absent in cold systems where the scatterers remain fixed during the measurement. A small redshift is still observed for dense clouds in Refs.\cite{pellegrino2014observation,jennewein2016} but could be specific to the geometry of the system.

Achieving large densities is concomitant with a vanishingly small transmission ${\cal T}$. It is therefore desirable to switch to a 2D or thin slab geometry in order to investigate the physical consequences of these resonant interaction effects at the macroscopic level. Using a 2D geometry also raises a fundamental question inspired by the monolayer semiconductor case: Can the light extinction through a plane of randomly positioned atoms be made arbitrarily large when increasing the atom density or does it remain finite, potentially introducing a maximum of light extinction through 2D random atomic samples independent of the atomic species of identical electronic spin?

In this article, we study the transmission of nearly resonant light through uniform slabs of atoms. We report experiments realized on a dense layer of atoms with a tunable density and thickness. For dense clouds, the transmission is strongly enhanced compared to the one expected from the single-atom response. We also observe a broadening and a blue shift of the resonance line on the order of the natural linewidth. This blue shift contrasts with the mean-field Lorentz-Lorenz red shift and is a signature of the strongly-correlated regime reached in our system because of dipole-dipole interactions \cite{Javanainen14}. To our knowledge, it is the first time that a blue shift is reported. In addition, we observe deviations of the resonance lineshape from the single-atom Lorentzian behavior, especially in the wings where the transmission decays more slowly.  We model this system with coupled dipole simulations complemented by a perturbative approach which qualitatively supports our observations.  After describing our experimental system in Sec.\,\ref{secII}, we investigate theoretically light scattering for the geometry explored in the experiment in Sec.\,\ref{secIII}. In Sec.\,\ref{secIV} we present our experimental results and compare them with theory. We conclude in Sec.\,\ref{secV}.


\section{Experimental methods}\label{secII}

\subsection{Cloud preparation}

We prepare a cloud of $^{87}$Rb atoms with typically $N=1.3(2)\times10^{5}$ atoms in the
$|F=1,m_{F}=-1\rangle$ state. The atoms are confined in an all-optical trap, described in more detail in \cite{Ville17}, with a strong harmonic confinement in the vertical direction $z$ with frequency $\omega_{z}/2\pi=2.3(2)\,{\rm kHz}$ leading to a gaussian density profile along this direction. The transverse confinement along the $x-$ and $y-$ directions is produced by a flat-bottom disk-shaped potential of diameter $2R=40\,\mu{\rm m}$. For our initial cloud temperature $\simeq 300$\,nK, there is no extended phase coherence in the cloud \footnote{The 2D phase-space density of the cloud is below the transverse condensation threshold as defined in \cite{Chomaz15}}. Taking into account this finite temperature, we compute for an ideal Bose gas an r.m.s. thickness $\Delta z=0.25(1)\,\mu$m, or equivalently $k\Delta z=2.0(1)$. This situation corresponds to $nk^{-2} \approx 1.5$, where $n=N/(\pi R^2)$ is the surface density and to a maximum density $\rho k^{-3} \approx 0.3$ at the trap center along $z$ where $\rho$ is the volume density. We tune the number of atoms that interact with light by partially transferring them to the $|F=2,m_{F}=-2\rangle$ state using a resonant microwave transition. Atoms in this state are sensitive to the probe excitation, contrary to the ones in the $|F=1,m_{F}=-1\rangle$ state. In this temperature range the Doppler broadening is about 3 orders of magnitude smaller than the natural linewidth of the atomic transition.

The cloud thickness is varied in a controlled way using mainly two techniques:
(i) Varying the vertical harmonic confinement by modifying the laser power in the blue-detuned lattice that traps the atoms, thus changing its frequency from $\omega_{z}/2\pi=1.1(2)\,{\rm kHz}$ to $\omega_{z}/2\pi=2.3(2)\,{\rm kHz}$. Using the ideal Bose gas statistics in the tight harmonic trap, for a gas of $N=1.3(2)\times10^{5}$ atoms at a temperature of $T\simeq 300\,$nK, this corresponds to r.m.s. thicknesses between 0.3\,$\mu$m and 0.6\,$\mu$m.
(ii) Allowing the atoms to expand for a short time after all traps have been switched off. The extent of the gas in the $xy$--direction does not vary significantly during the time of flight (ToF) (duration between 0.7\,ms and 4.7\,ms). In that case, the r.m.s. thickness varies between 3\,$\mu$m and  25\,$\mu$m. For the densest clouds, the thickness is also influenced by the measurement itself. Indeed, the light-induced dipole-dipole forces between atoms lead to an increase of the size of the cloud during the probing. In the densest case, we estimate from measurements of the velocity distribution after an excitation with a duration of $\tau=10\,\mu$s that the thickness averaged over the pulse duration is increased by $\sim 20\,\%$. In some experiments, in which the signal is large enough, we limit this effect by reducing the probe duration $\tau$ to 3\,$\mu$s.

\begin{figure}[hbt!!!]
\begin{centering}
\includegraphics[width=8.6cm]{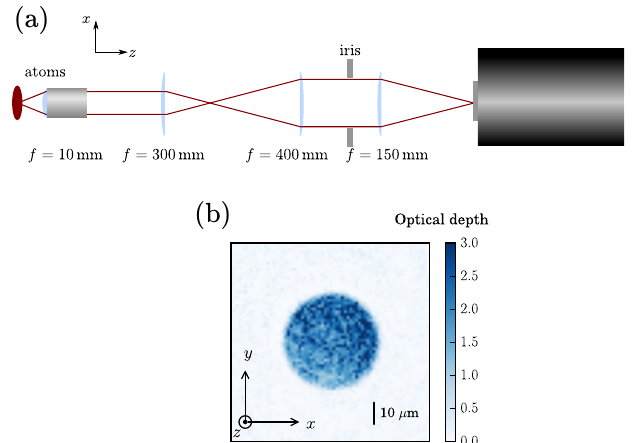} 
\par\end{centering}

\caption{(a) Schematic representation of the imaging setup. The atoms are confined by a single, disk-shaped potential which is imaged using a microscope objective onto a back-illuminated CCD camera. The numerical aperture of the system is limited to $\approx 0.2$ using an iris in the Fourier plane of the atoms to limit the collected fluorescence light. (b) Typical in-situ image obtained on a back illuminated CCD camera of the in-plane density distribution averaged over three individual measurements. For this example, the atom surface density is $n=25\,\mu$m$^{-2}$. We extract a region of interest with uniform density for our analysis with a typical area of 200\,$\mu$m$^2$. \label{fig1}}
\end{figure}

\subsection{Transmission measurement}

We probe the response of the cloud by measuring the transmission of a laser beam propagating along the $z-$direction (See Fig.\,1). The light is linearly polarized along the $x-$axis and tuned close to the $|F=2\rangle\rightarrow|F'=3\rangle$ D$_2$ transition. The duration of the light pulse is fixed to  $10\,\mu$s for most experiments and we limit the imaging intensity $I$ to the weakly saturating regime with $0.075<I/I_{\rm sat}<0.2$, where $I_{{\rm sat}}\simeq 1.67\,$mW/cm$^2$ is the resonant saturation intensity. We define $\Delta\nu$ as the detuning of the laser beam with respect to the single-atom resonance. The cloud intensity transmission $|\mathcal{T}|^2$ is extracted by comparing images with and without atoms and we compute the optical depth ${\rm D}=-\ln |\mathcal{T}|^2$ (see Sec.\,\ref{secII}.C). The numerical aperture of the optical system is limited on purpose to minimize the collection of fluorescence light from directions different from the propagation direction of the light beam. 

\begin{figure}[!hbt]
\begin{centering}
\includegraphics[width=6.6cm]{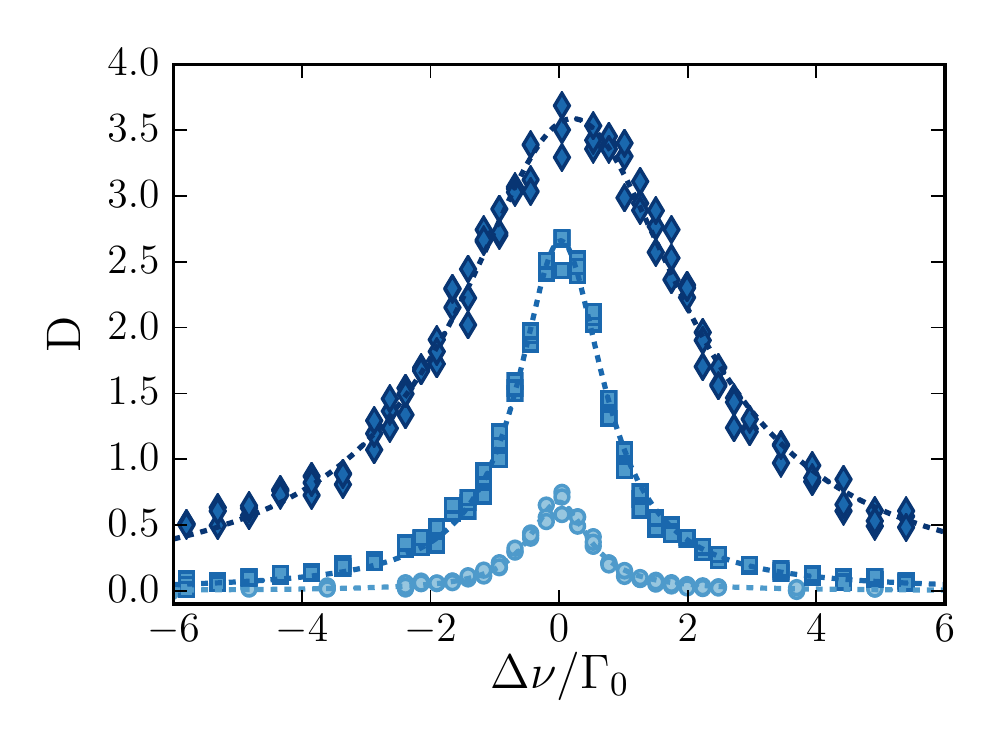} 
\par\end{centering}
\caption{Example of resonance curves. Symbols represent the experimental data, and the corresponding dashed lines are Lorentzian fits. All curves are taken with the cloud thickness $k\Delta z=2.4(6)$ and for surface densities of $nk^{-2}=0.06(1)$ (circles), $0.38(6)$ (squares) and $1.5(2)$ (diamonds). The errors on the fitted parameters are determined using a basic bootstrap analysis, repeating the fitting procedure 100 times on a set of random points drawn from the original set of data, of the same length as this original set.\label{fig2}}
\end{figure}

\subsection{Computation of the optical depth}
We extract the optical depth (${\rm D}$) of the clouds by comparing pictures with and without atoms. The read-out noise on the count number $N_{\rm count}$ is ${\rm d}N_{\rm count}\sim 5$ per pixel. We subtract from these images equivalent pictures without any imaging pulse to remove the background counts and obtain two pictures $M_{\rm with}$ and $M_{\rm without}$. The typical noise on the count number per pixel is thus ${\rm d}N=\sqrt{2}\,{\rm d}N_{\rm count}\sim 7$.

The magnification of the optical system is 11.25, leading to an effective pixel size in the plane of the atoms of 1.16\,$\mu$m. The typical mean number of counts per pixel accumulated during the $10\,\mu$s imaging pulse is 80 on the picture without atoms. We optimize the signal-to-noise ratio by summing all the pixels in the region of interest for $M_{\rm with}$ and $M_{\rm without}$. 
This yields a total count number in the picture with atoms $N_{\rm with}$ and without atoms $N_{\rm without}$ from which we compute the optical depth: ${\rm D} = -\ln(N_{\rm with}/N_{\rm without})$. The region of interest varies with the time-of-flight of the cloud. This region is a disk that ensures that we consider a part of the cloud with approximately constant density (with  15\% rms fluctuations), comprising typically 200 pixels. With these imaging parameters we can reliably measure optical depths up to
4 but we conservatively fit only data for which ${\rm D < 3}$. At low densities, the statistical error on ${\rm D}$ due to the read-out noise is about $0.01$.  At ${\rm D \sim 3}$, it reaches 0.12.

\subsection{Atom number calibration}
As demonstrated in this article, dipole-dipole interactions strongly modify the response of the atomic cloud to resonant light and make an atom number calibration difficult. In this work, we measure the atom number with absorption imaging for different amounts of atoms transferred by a coherent microwave field from the $|F=1,m_{F}=-1\rangle$ ``dark" state to the $|F=2,m_{F}=-2\rangle$ state in which the atoms are resonant with the linearly polarized probe light. We perform resonant Rabi oscillations for this coherent transfer and fit the measured atom number  as a function of time by a sinus square function. We select points with an optical depth below 1, to limit the influence of dipole-dipole interactions. This corresponds to small microwave pulse area or to an area close to a $2\pi$ pulse, to make the fit more robust. From the measured optical depth ${\rm D}$, we extract $n k^{-2}=(15/7)\,{\rm D}/(6\pi)$. The factor 7/15 corresponds to the average of equally-weighted squared Clebsch-Gordan coefficients for linearly polarized light resonant with the $F=2$ to $F'=3$ transition. This model does not take into account possible optical pumping effects that could lead to an unequal contribution from the different transitions and hence a systematic error on the determination of the atom number.

\subsection{Experimental protocol}
Our basic transmission measurements consist in scanning the detuning $\Delta\nu$ close to the $F=2$ to $F'=3$ resonance ($|\Delta\nu|< 30\,$MHz) and in measuring the optical depth at a fixed density. The other hyperfine levels $F'=2,1,0$ of the excited $5P_{3/2}$ level play a negligible role for this detuning range. The position of the single-atom resonance is independently calibrated using a dilute cloud. The precision on this calibration is of $0.03\,\Gamma_0$, where $\Gamma_0/2\pi=6.1\,$\,MHz is the atomic linewidth.  The measured resonance curves are fitted with a Lorentzian function:
\begin{equation}
\Delta\nu\mapsto {\rm D_{max}}/[1+4\left(\Delta\nu-\nu_{0}\right)^{2}/\Gamma^2].
\end{equation}
This function captures well the central shape of the curve for thin gases, as seen in the examples of Fig.\,\ref{fig2}. When increasing the atomic density we observe a broadening of the line $\Gamma>\Gamma_0$, a non-linear increase of the maximal optical depth ${\rm D}_{\rm max}$ and a blue shift $\nu_{0}>0$. In Sec. \ref{secIV} we present the evolution of these fitted parameters for different densities and thicknesses.
Note that in our analysis all points with values of {\rm D} above 3 are discarded to avoid potential systematic errors. Whereas this threshold has little influence for thin clouds (as shown in Fig. \ref{fig2}) for which the maximal optical depths are not large compared to the threshold, for thick gases this typically removes the measurements at detunings smaller than 1.5\,$\Gamma_0$. Hence, in this case, we consider the amplitude and the width of the fits to be not reliable and we use the position of the maximum of the resonance $\nu_0$ with caution.

We investigate the dependence of the fit parameters  ${\rm D}_{\rm max}$, $\Gamma$ and $\nu_{0}$ for different atomic clouds in Sec\,\ref{secIV}. These results are compared to the prediction from a theoretical model that we describe in the following Section.


\section{Theoretical description}\label{secIII}
Light scattering by a dense sample of emitters is a complex many-body problem and it is quite challenging to describe. The slab geometry is a textbook situation which has been largely explored. A recent detailed study of the slab geometry can be found in Ref.\,\cite{Javanainen17}. We focus in this section first on a perturbative approach which is valid for low enough densities. We then report coupled dipole simulations following the method presented in \cite{Chomaz12} but extended with a finite-size scaling approach to address the situation of large slabs. We also discuss the regime of validity for these two approaches.

\subsection{Perturbative approach}
We describe here a semi-analytical model accounting for the multiple scattering of light by a dilute atom sample, inspired from reference \cite{morice1995refractive}. By taking into account multiple scattering processes between atom pairs, it provides the first correction to the Beer-Lambert law when decreasing the mean distance $l$ between nearest neighbors towards $k^{-3}$.

\subsubsection{Index of refraction of a homogeneous system}

\begin{figure}[hbt!!]
\includegraphics[width=6.6cm]{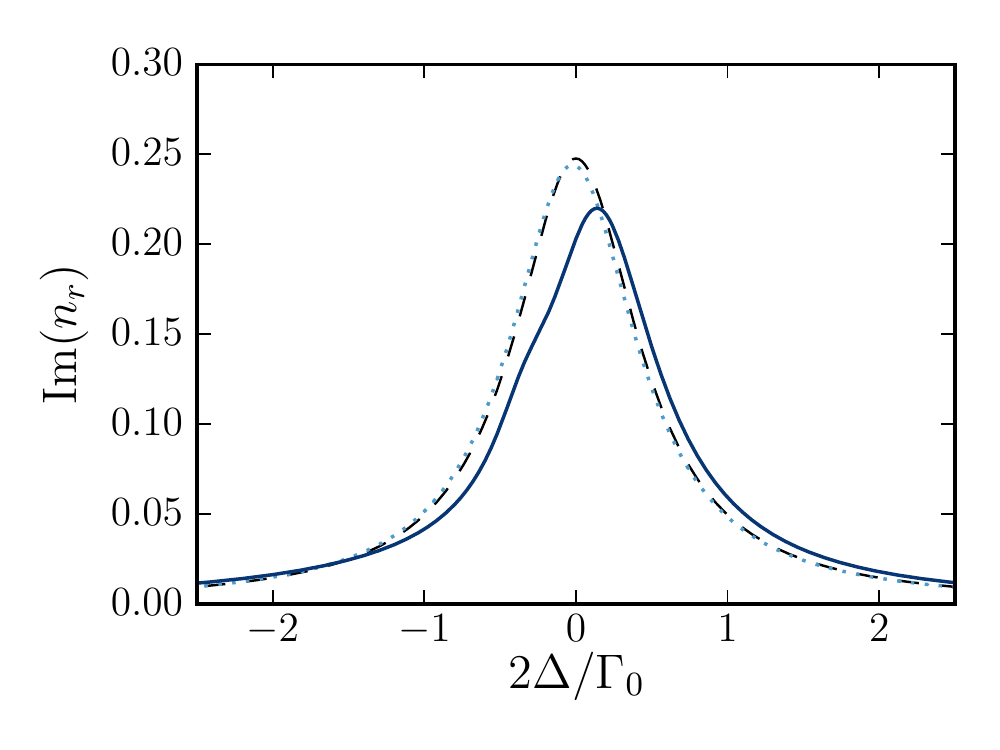}
\caption{\label{fig3}
Imaginary part of the index of refraction of an homogeneous atomic sample of density $\rho k^{-3}\simeq 0.026$. The three curves corresponds to the absorption of independent atoms (dashed black line), to the resonance line taking into account the Lorentz-Lorenz correction (dotted blue line), and to the perturbative analysis discussed in the text (solid black line), which takes into account multiple scattering of photons between pairs of atoms \cite{morice1995refractive}.}
\end{figure}

In reference \cite{morice1995refractive}, the index of refraction of a homogeneous dilute atomic gas was calculated, taking into account the first non-linear effects occurring when increasing the volume atom density. The small parameter governing the perturbative expansion is $\rho k^{-3}$, where $\rho$ is the atom density. At second  order in $\rho k^{-3}$ two physical effects contribute to the refraction index, namely the effect of the quantum statistics of atoms on their position distribution, and the dipole-dipole interactions occurring between nearby atoms after one photon absorption. Here we expect the effect of quantum statistics to remain small, and thus neglect it hereafter (see Ref.\,\cite{Bons16} for a recent measurement of this effect). Including the effect of multiple scattering processes between atom pairs, one obtains the following expression for the refractive index: 
\begin{align}
n_r&=1+\frac{\alpha\rho}{1-\alpha\rho/3+\beta\rho}\label{eq_index}\\
\beta&=
-\int\dd\rr\left[\frac{\alpha^2\GG'^2+\alpha^3\GG'^3\mathrm{e}^{-ikz}}{1-\alpha^2\GG'^2}\right]_{xx}(\rr)
\end{align}
where we introduced the atom polarisability $\alpha=6\pi ik^{-3}/(1-2i\delta/\Gamma)$ and the Green function $[\GG]$ of an oscillating dipole
\begin{align}
\GG_{\alpha\beta}(\rr)=&-\frac{1}{3}\delta(\rr)\delta_{\alpha\beta}+\GG'_{\alpha\beta}(\rr),\nonumber\\
\GG'_{\alpha\beta}(\rr)=&-\frac{k^3}{4\pi }\frac{e^{ikr}}{kr}\bigg[\left(1+\frac{3i}{kr}-\frac{3}{(kr)^2}\right)\frac{r_{\alpha}r_{\beta}}{r^2}\nonumber\\
&\phantom{-\frac{1}{4\pi k^3}\frac{e^{ikr}}{kr}\bigg[}-\left(1+\frac{i}{kr}-\frac{1}{(kr)^2}\right)\delta_{\alpha\beta}\bigg],
\end{align}
in which retardation effects are neglected \cite{Milonni74}. Note that for a thermal atomic sample of Doppler width larger than $\Gamma$, we expect an averaging of the coherent term $\beta$ to zero due to the random Doppler shifts. When setting $\beta=0$ in Eq.\,\ref{eq_index} we recover the common Lorentz-Lorenz shift of the atomic resonance \cite{ruostekoski1997lorentz}. We plot in Fig.\,\ref{fig3} the imaginary part of the index of refraction as a function of the detuning $\delta$, for a typical atom density used in the experiment (solid line) and compare with the single-atom response with (dotted line) and without (dashed line) Lorentz-Lorenz correction. The resonance line is modified by dipole-dipole interactions and we observe a blue shift of the position of the maximum of the resonance \cite{cherroret2016induced}.

\subsubsection{Transmission through an infinite slab with a gaussian density profile}

In order to account more precisely for the light absorption occurring in the experiment, we extend the perturbative analysis of light scattering to inhomogeneous atom distributions, for which the notion of index of refraction may not be well-defined.
The atom distribution  is modeled by an average density distribution $\rho(z)$ of infinite extent along $x$ and $y$, and depending on $z$ only, as $\rho(z)=\rho_0\exp[-z^2/(2\Delta z^2)]$.
We describe the propagation of light along $z$ in the atomic sample. The incoming electric field is denoted as $E_0e^{i(kz-\omega t)}\ex$. 
The total electric field, written as $\mathbf{E}(z)e^{-i\omega t} $, is given by the sum of the incoming field and the field radiated by the excited atomic dipoles:
\begin{equation}
\mathbf{E}(z)=E_0e^{ikz}\ex+\int\dd^3\rr'\,\rho(z')\frac{ [\GG(\rr-\rr')]}{\epsilon_0}\DD(z'),
\end{equation}
where  $\DD(z)$ is the dipole amplitude of an atom located at $z$ and $\epsilon_0$ is the vacuum permittivity. The integral over $x$ and $y$ can be performed analytically, leading to the expression
\begin{equation}\label{eq_E}
\mathbf{E}(z)=E_0e^{ikz}\ex+\frac{i k}{2\epsilon_0}\int\dd z'\,\rho(z')e^{ik|z-z'|}\DD_\perp(z'),
\end{equation}
where $\DD_\perp(z)$ is the dipole amplitude  projected in the $x,y$ plane.

The dipole amplitude can be calculated from the atom polarisability $\alpha$ and the electric field at the atom position.
Taking into account multiple light scattering between atom pairs, we obtain a self-consistent expression for the dipole amplitude, valid up to first order in atom density, as $\DD(z)=d(z)\ex$, with
\begin{align}
d(z)&=\alpha\epsilon_0E_0e^{ikz}\nonumber\\
&+
\int\dd\rr'\,\rho(z')\bigg\{\left[\frac{\alpha\GG}{1-\alpha^2\GG^2}\right]_{xx}\!\!\!\!\!\!(\rr-\rr')d(z')\nonumber\\
&
\phantom{+\int\dd\rr'\,\rho(z')\bigg\{}
+\left[\frac{\alpha^2\GG^2}{1-\alpha^2\GG^2}\right]_{xx}\!\!\!\!\!\!(\rr-\rr')d(z)\bigg\}.\label{eq_d}
\end{align}
Note that the dipole amplitude also features a component along $z$, but it would appear in the perturbative expansion in the atom density at higher orders. 

The electric field and dipole amplitude are numerically computed by solving the linear system (\ref{eq_E})-(\ref{eq_d}). The optical depth is then calculated as $\mathrm{D}=-\ln(|E(z)|^2/|E_0|^2)$ for $z\gg \Delta z$. The results of this approach will be displayed and quantitatively compared to coupled dipole simulations in the next subsection.

\subsection{Coupled dipole simulations}

\subsubsection{Methods}

Our second approach to simulate the experiments follows the description in Ref. \cite{Chomaz12} and uses a coupled dipole model. We consider atoms with a $J=0$ to $J=1$ transition. For a given surface density $n$ and thickness $\Delta z$ we draw the positions of the $N$ atoms with a uniform distribution in the $xy$ plane and a Gaussian distribution along the $z$ direction. The number of atoms and hence the disk radius is varied to perform finite-size scaling. For a given detuning and a linear polarization along $x$ of the incoming field, we compute the steady-state value of each dipole $\bm{d}_j$ which is induced by the sum of the contributions from the laser field and from all the other dipoles in the system. The second contribution is obtained thanks to the tensor Green function $\GG$ giving the field radiated at position $\mathbf r$ by a dipole located at origin.

Practically, the values of the $N$ dipoles are obtained by numerically solving a set of $3N$ linear equations, which limits the atom number to a few thousands, a much lower value than in the experiment (where we have up to $10^5$ atoms). From the values of the dipoles we obtain the transmission $\mathcal{T}$ of the sample:
\begin{eqnarray}
\mathcal{T}=1-\frac{i}{2} \sigma \frac{ n k^{-2}}{N} \sum_j \frac{k^3}{6\pi \epsilon_0 E_L} d_{j,x} e^{-ikz_j} 
\end{eqnarray}
where $z_j$ is the vertical coordinate of the $j$-th atom, $E_L$ the incoming electric field, and $d_{j,x}$ is the $x$ component of the dipole of the $j$-th atom. From the transmission, we compute the optical depth $\mathrm{D}=-\ln |\mathcal T|^2$ and fit the resonance line with a Lorentzian line shape to extract, as for the experimental results, the maximum, the position and the width of the line.

\begin{figure}[hbt!!!]
\includegraphics[width=8.6cm]{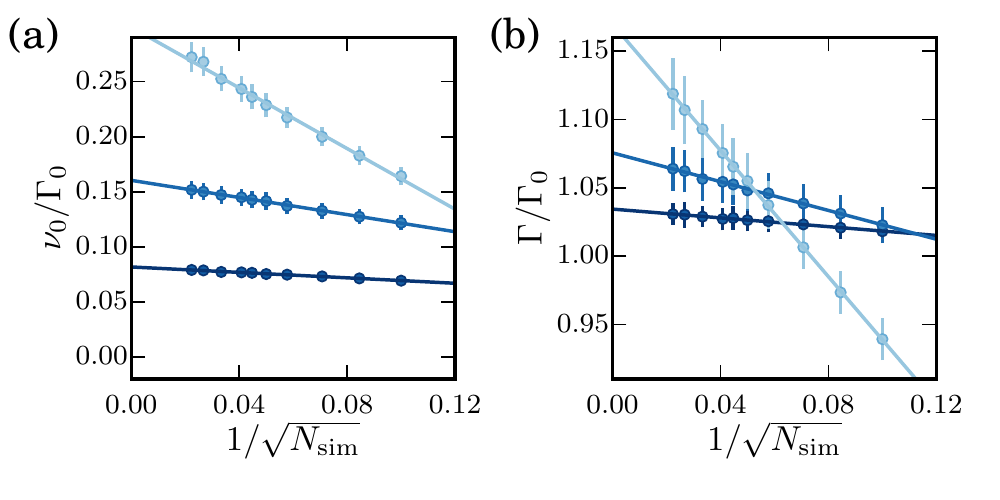}
\caption{Example of finite-size scaling to determine (a) the position of the maximum of the resonance $ \nu_0$ and (b) the width of the resonance. Here $k \Delta z=1.6$ and (from bottom to top) $ n k^{-2}=0.05$, 0.11 and 0.21. Simulations are repeated for different atom number $N_{\rm sim}$. The number of averages ranges from 75 (left points, $N_{\rm sim}=2000$) to 25\,000 (right points, $N_{\rm sim}=100$).  When plotting the shift as a function of $1/\sqrt{N_{\rm sim}} \propto 1/R$, and for low enough densities, data points are aligned and allow for a finite-size scaling. Vertical error bars represent the standard error obtained when averaging the results over many random atomic distributions.   \label{fig4}}
\end{figure}

\begin{figure}[hbt!!!]
\includegraphics[width=8.6cm]{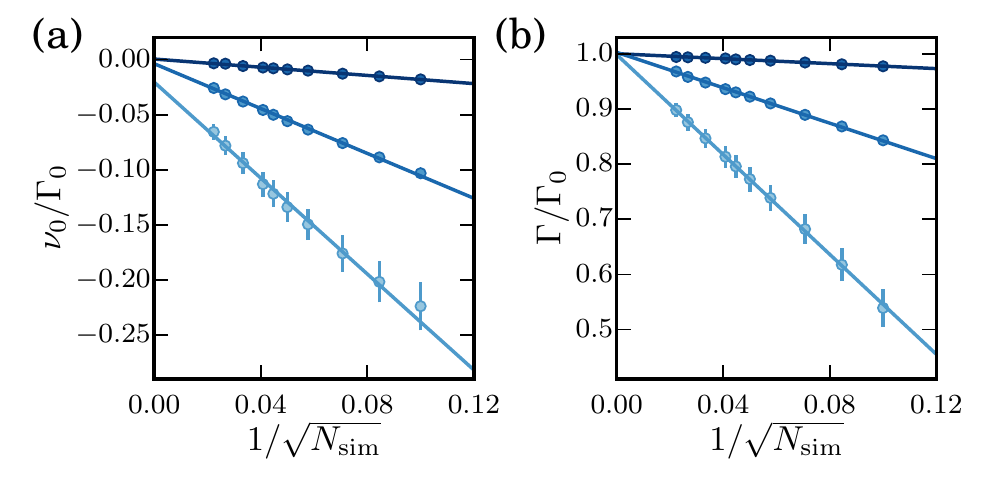}
\caption{Example of finite-size scaling to determine (a) the position of the maximum of the resonance $\nu_0$ and (b) the width $ \Gamma$ of the resonance. Here, $k \Delta z=80$ and (from top to bottom) $ n k^{-2}=0.027$, 0.08 and 0.13.  Simulations are repeated for different atom number $N_{\rm sim}$. The number of averages ranges from 75 (left points, $N_{\rm sim}=2000$) to 25\,000 (right points, $N_{\rm sim}=100$). Vertical error bars represent the standard error obtained when averaging the result over many random atomic distributions.  \label{fig5}}
\end{figure}

\begin{figure*}[hbt!!!]
\includegraphics[width=\textwidth]{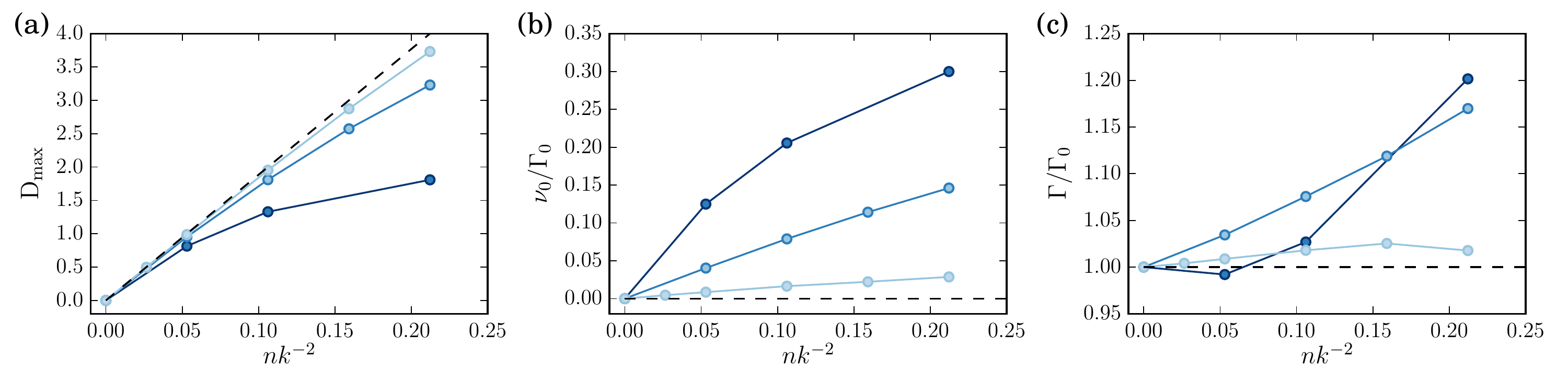}
\caption{Coupled dipole simulations for different thicknesses. (a) Maximal optical depth, (b) Position of the maximum of the line, (c) Width of the resonance line. We report results for $k\Delta z=$0, 1.6 and 8, the darkest lines corresponding to the smallest thicknesses. The black dashed lines correspond to the single-atom response. \label{fig6}}
\end{figure*}

As the number of atoms used in the simulations is limited, it is important to verify the result of the simulations is independent of the atom number. In this work, we are mostly interested in the response of an infinitely large system in the $xy-$plane. It is indeed the situation considered in the perturbative approach and in the experimental system for which the diameter is larger than $ 300\, k^{-1}$ and where finite-size effects should be small. The atom number in the simulations is typically two order of magnitudes lower than in the experiment and finite-size effects could become important. For instance, some diffraction effects due to the sharp edge of the disk could play a role \cite{Javanainen17}. Consequently, we varied the atom number in the simulations and observed, for simulated clouds with small radii, a significant dependence of the simulation results on the atom number. We have developed a finite-size scaling approach to circumvent this limitation. We focus in the following on transmission measurements as in the experiment.

We show two examples of this finite-size scaling approach for $k \Delta z=1.6$ in Fig.\,\ref{fig4} and  $k \Delta  z=80$ in Fig.\,\ref{fig5}. For low enough surface densities, the results of the simulations (maximal optical depth, width, shift,...) for different atom numbers in the simulation are aligned, when plotted as a function of $1/\sqrt{N_{\rm sim}}$, and allow for the desired finite-size scaling.  All the results presented in this section and in Sec.\,\ref{secIV} \footnote{except for Fig.\,\ref{fig10}(b) for which simulations are performed for a fixed atom number of 2000.} are obtained by taking the extrapolation to an infinite system size, which corresponds to the offset of the linear fit in Figs.\,\ref{fig4} and  \ref{fig5}.

Interestingly, we observe in Fig.\,\ref{fig4} for a thin cloud that considering a finite-size system only leads to a small underestimate of the blue shift of the resonance. However, for thicker slabs, such as in Fig.\,\ref{fig5}, we get, for finite systems, a small red shift and a narrowing of the line. Considering our experimental system, we have $1/\sqrt{N} \approx 0.003$, leading to a small correction according to the fits in Fig.\,\ref{fig5}. However, for such thick systems we are able to simulate only systems with low $nk^{-2}$, typically 0.1, whereas we can reach densities 15 times larger in the experiment, which could enhance finite-size effects. Simulation of thick and optically dense slabs is thus challenging and the crossover between the thin slab situation explored in this article and the thick regime is an interesting perspective of this work.


\subsubsection{Role of the thickness and density of the cloud}
We now investigate the results of coupled dipole simulations for different densities and thicknesses of the atomic cloud. We limit the study to low densities, for which the finite-size scaling approach works. It is important to note that the computed line shapes deviate significantly from a Lorentzian shape and become asymmetric. Consequently there is not a unique definition for the center of the line and for its width. In our analysis, we fit the resonance lines around their maximum with a typical range of $\pm 0.5\,\Gamma$. The shift thus corresponds to a variation of the position of the maximum of the line and the ``width" characterizes the curvature of the line around its maximum. The results of these fits are reported in Fig.\,\ref{fig6} as a function of surface density for different thicknesses. In these plots, we observe the same features as qualitatively described in Sec.\,\ref{secII}: A decrease of the maximal optical depth with respect to the single-atom response (Fig.\,\ref{fig6}(a)), a blue shift of the position of the maximum (Fig.\,\ref{fig6}(b)) and a broadening of the line  (Fig.\,\ref{fig6}(c)). For a fixed thickness, these effects increase with surface density and for a fixed surface density they are more pronounced for lower thicknesses. Note that we only explore here surface densities lower than 0.25 which is quite lower than the maximum experimental value ($\sim 1.5)$. Whereas our finite-size scaling approach can be well-extended for very thin systems ($k \Delta z <1$) it fails for thick and optically dense systems \footnote{We guess that this is due to the limited atom number used in the simulation which prevents from investigating a regime where the geometry of the simulated cloud has an aspect ratio similar to the experimental one.}.

\begin{figure}[hbt!!!]
\includegraphics[width=6.6cm]{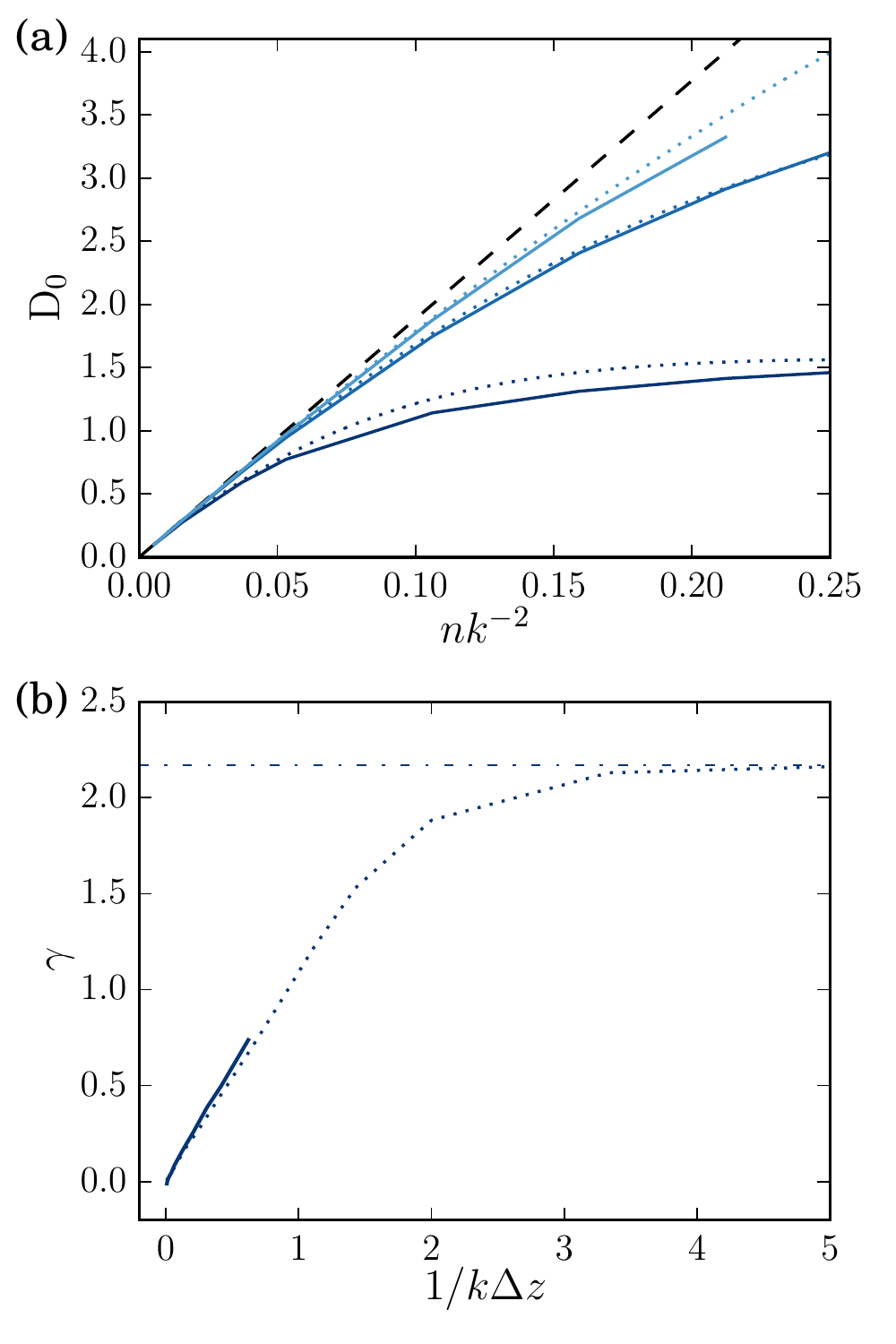}
\caption{ Comparison between the coupled dipole simulations and the perturbative model. (a) Behavior of the optical depth at the single-atom resonance $\mathcal D_0$  with surface density for different thicknesses ($k\Delta z =$0, 1.6 and 3.2, from bottom to top). Coupled dipole simulations are shown as solid lines, perturbative approach as dotted lines and the dashed line is the Beer-Lambert prediction. (b) Slope $\gamma$ of the blue shift $\nu_0=\gamma n k^{-2}$ as a function of the inverse thickness $1/k\Delta z$. The solid line is the result of the coupled dipole model, the dash-dotted line is the zero-thickness coupled dipole result ($1/k \Delta z \rightarrow \infty$), the dotted line is the perturbative model.\label{fig7}}
\end{figure}

\subsubsection{Comparison with the perturbative model}
The perturbative approach is limited to low densities $\rho k^{-3} \ll 1$ but it gives the response of an infinitely expanded cloud in the transverse direction. Coupled dipole simulations can in principle address arbitrarily large densities but the number of atoms considered in a simulation is limited, and thus for a given density the size of the system is limited. Coupled dipole simulations are thus more relevant for thin and dense sample and the perturbative approach more suited for non-zero thickness samples.

In Fig. \ref{fig7} we choose two illustrative examples to confirm, in the regime where both models could be used, that these two approaches are in quantitative agreement. In Fig.\,\ref{fig7}(a) we compare the maximum optical depth as a function of surface density for three different thicknesses. The perturbative approach is typically valid, for this set of thicknesses, up to $nk^{-2} \sim 0.1$. We investigate the shift of the position of the maximum in Fig.\,\ref{fig7}(b). We report, as a function of the inverse thickness ($1/k \Delta z$), the slope $\gamma$ of the shift with density, $\nu_0=\gamma n k^{-2}$, computed for surface densities below 0.1. The dotted line is the result from the perturbative approach, the solid line corresponds to coupled dipole simulations and the dash-dotted line to the result for zero thickness. The perturbative approach approximates well coupled dipole simulations. This result also confirms that the finite-size scaling approach provides a good determination of the response of an infinite system in the $xy-$direction.

We have identified in this Section the specific features of the transmission of light trough a dense slab of atoms. We focus here on the transmission coefficient to show that we observe the same features in the experiment and we will make a quantitative comparison between our experimental findings and the results obtained with coupled dipole simulations.  Our theoretical analysis is complemented by a study of the reflection coefficient of a strictly 2D gas detailed in Appendix A.

\begin{figure}[!hbt]
\begin{centering}
\includegraphics[width=7cm]{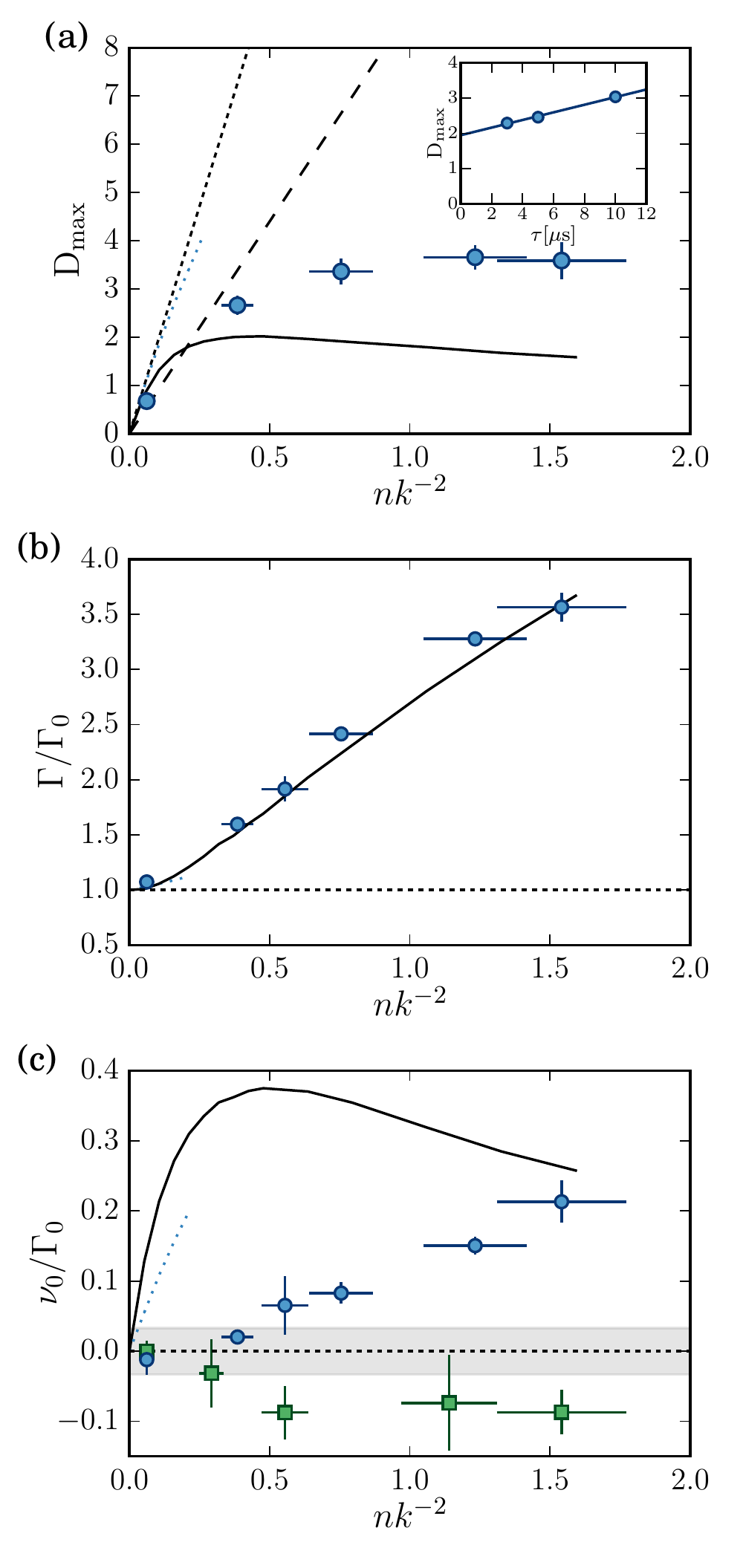} 
\par\end{centering}

\caption{Maximum optical depth (a), broadening (b) and frequency shift (c) of the resonance line for our thinnest samples with $k\Delta z = 2.4(6)$  (circles) and for thicker samples with $k\Delta z = 30(8)$ (squares). In (a) the shaded area represents the uncertainty in the frequency calibration of the single-atom resonance. In (a) and (b), the dark black solid (resp. light blue dotted) line is the prediction of the coupled dipole model for $k \Delta z=0$ (resp. $k \Delta z=2.4$) in its accessible range of densities. The dashed lines represent the single-atom response. \label{fig9} }
\end{figure}


\section{Experimental results}\label{secIV}

We show in Fig.\,\ref{fig9} the results of the experiments introduced in Sec.\,\ref{secII}. The fitted ${\rm D_{max}}$ for different surface densities is shown in Fig.\,\ref{fig9}(a). We compare these results to the Beer-Lambert prediction (narrow dashes) ${\rm D_{BL}}=n\sigma_0$ and to the same prediction corrected by a factor 7/15 (large dashes). This factor is the average of the Clebsch-Gordan coefficients relevant for $\pi$-polarized light tuned close to the $|F=2\rangle\rightarrow|F'=3\rangle$ transition and, as discussed previously, it is included in the calibration of the atom number. At large surface densities, we observe an important deviation from this corrected Beer-Lambert prediction: we measure that ${\rm D_{max}}$ seems to saturate around  ${\rm D_{max}}\approx 3.5$ whereas ${\rm D_{BL}}\approx 13$ \footnote{The observed saturation is not due to our detection procedure. In the fitting procedure, we only select points with a measured optical depth below 3 to avoid any bias.}. We also show the prediction of the coupled dipole model, as a solid line for the full range of surface densities at $k\Delta z=0$ and as a dotted line for the numerically accessible range of surface densities at $k\Delta z=2.4$. The coupled dipole simulation at $k\Delta z=0$ shows the same trend as in the experiment  but with ${\rm D_{max}}$ now bounded by 2. 
A reason could be the non-zero thickness of the atomic slab. In order to test this hypothesis, we investigated the influence of probing duration for the largest density. For such a density we could decrease the pulse duration while keeping a good enough signal to noise ratio (see inset in Fig. 2(b)). For a  shorter probing duration, hence for a smaller expansion of the cloud, ${\rm D_{max}}$ decreases, in qualitative agreement with the expected effect of the finite thickness.

The saturation of the optical depth with density is a counterintuitive feature. It shows that increasing the surface density of an atomic layer does not lead to an increase of its optical depth. Coupled dipole simulations at $k\Delta z=0$ even show that the system becomes slightly more transparent as the surface density is increased. These behavior may be explained qualitatively by the broadening of the distribution of resonance frequencies of the eigenmodes of this many-body system. A dense system scatters light for a large range of detunings but the cross section at a given detuning saturates or becomes lower as the surface density is increased.

We display in Fig. \ref{fig9}(b) the width $\Gamma$ of the Lorentzian fits for $k\Delta z=2.4(6)$ along with coupled dipole simulation results \cite{Note3}. We observe a broadening of the resonance line up to more than $3\,\Gamma_0$. This broadening is confirmed by the simulation results for $k\Delta z=0$ (solid line). Note that the exact agreement with the experimental data should be considered as coincidental. The range on which we can compute the broadening for $k\Delta z=2.4$ (light dotted line) is too small to discuss a possible agreement.

We show in Fig.\,\ref{fig9}(c) the evolution of $\nu_0$ with density. A blue shift, reaching $0.2\,\Gamma_0$ for the largest density, is observed. At the largest density, an even larger shift is observed when decreasing the pulse duration ($\approx 0.4\,\Gamma_0$, not shown here). We also display the result of the coupled dipole model for the cases $k\Delta z=2.4$ and $k\Delta z=0$. Both simulations confirm the blue shift but predict a different behavior and a larger effect. In addition, we show the variation of $\nu_0$ for a thick cloud with $k\Delta z=30(8)$. In that case we observe a marginally significant red shift \footnote{For thick clouds the expected optical depth of the cloud becomes larger than the maximal value we can detect. We then only fit the points with a measured optical depth below 3. From this fit we cannot estimate reliably the value of the maximum and width of the resonance line, but we can get an estimate of the ``center'' of the line which may deviate from the position of the maximum of the resonance for an asymmetric lineshape.}.

\begin{figure}[!hbt]
\begin{centering}
\includegraphics[width=8.6cm]{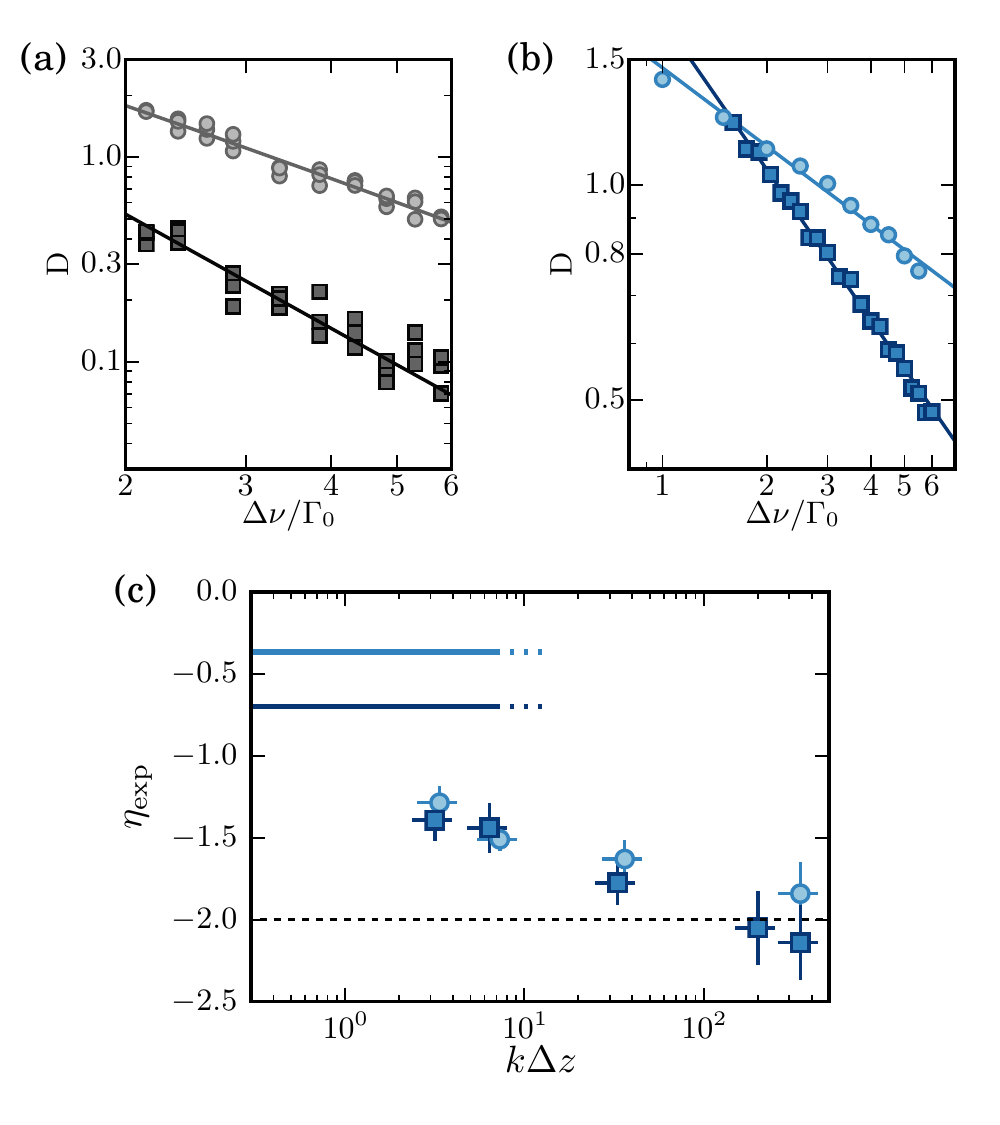}
\par\end{centering}
\caption{Non-Lorentzian wings of the resonance line. (a) Two examples of the scaling of optical depth with $\Delta \nu $ (blue side), in log-log scale, for $k \Delta z=2.4(6)$ (circles) and for $k\Delta  z=350(90)$ (squares) and their power-law fit. (b) Coupled dipole simulations at  zero thickness and for $n k^{-2}=1.5(2)$ . The optical depth is plotted as a function of detuning (resp. minus the detuning) from the resonance line, for the blue (circles) (resp. red (squares)) side. The solid lines are power-law fits. (c) Experimental results for $n k^{-2}=1.5(2)$. Circles (resp. squares) represent the fitted exponents $\eta_{{\rm r}}$ (resp. $\eta_{{\rm b}}$) to the far-detuned regions of the resonance line on the red and blue side, respectively. The fit function is $\Delta\nu\mapsto {\rm D}(\Delta\nu)=A\,\left(\Delta\nu-\nu_0\right)^{\eta}$. The error on the fitted exponents is also determined using a bootstrap analysis. The horizontal dashed black line ($\eta=-2$) emphasizes the expected asymptotic value for low densities for a Lorentzian line (at large detunings).\label{fig10}}
\end{figure}

The experimental observation of a blue shift has never been reported experimentally to our knowledge. It is in stark contrast, both in amplitude and in sign, with the mean-field prediction of the Lorentz-Lorenz red shift $ \nu_0^{{\rm MF}}/\Gamma_0=-\pi \rho k^{-3}=-\sqrt{\pi/2}\, n k^{-2}/(k\Delta z)$, written here at the center of the cloud along $z$. The failure of the Lorentz-Lorenz prediction for cold atom systems has already been observed and discussed for instance in Refs \citep{Javanainen14,Bromley16,jennewein2016}. As discussed with the perturbative approach in Sec.\,\ref{secIII}, the Lorentz-Lorenz contribution is still present but it is (over)compensated by multiple scattering effects for a set of fixed scatterers. In hot vapors, where the Doppler effect is large, the contribution of multiple scattering vanishes and thus the Lorentz-Lorenz contribution alone is observed. The related Cooperative Lamb shift has been recently demonstrated in hot vapor of atoms confined in a thin slab in Ref.\,\cite{Keaveney2012}. In the cold regime where scatterers are fixed, such effects are not expected \cite{Javanainen17}. However, in these recent studies with dense and cold samples a small red shift is still observed \citep{Javanainen14,Bromley16,jennewein2016}. This difference on the sign of the frequency shift with respect to the results obtained in this work may be explained by residual inhomogeneous broadening induced by the finite temperature or the diluteness of the sample in Ref. \citep{Bromley16} and by the specific geometry in Ref. \cite{jennewein2016}, where the size of the atomic cloud is comparable to $\lambda$ and where diffraction effects may play an important role. As discussed in Sec.\,\ref{secIII}, our observation of a blue shift is a general result which applies to the infinite slab. It is robust to a wide range of thicknesses and density, and while we computed it theoretically for a two-level system, it also shows up experimentally in a more complex atomic level structure. It was also predicted  in Ref. \cite{Javanainen17} but for a uniform distribution along the $z$ axis instead of the Gaussian profile considered in this work, and also discussed in \cite{cherroret2016induced}. Consequently, we believe that it is an important and generic feature of light scattering in a extended cloud of fixed randomly distributed scatterers.

Finally, we compare the lineshape of the resonance with the Lorentzian shape expected for a single atom. We measure for $n k^{-2}=1.5(2)$, the optical depth at large detunings, and for various cloud thicknesses. We fit it with a power law on the red-detuned (resp. blue-detuned) frequency interval with exponent $\eta_{{\rm r}}$ (resp. $\eta_{{\rm b}}$) as shown, for two examples, in  Fig.\,\ref{fig10}(a). If the behavior were indeed Lorentzian, the exponents should be  $-2$ in the limit of large detuning. As seen in Fig.\,\ref{fig4}(c), for the thinnest gases, the fitted exponents are significantly different from the expected value and can reach values up to $-1.3$, showing the strong influence of dipole-dipole interactions in our system. We show the result of coupled dipole simulations  for $k \Delta z=0$ in Fig.\,\ref{fig4}(b) along with their power-law fit. We extract the exponents $\eta_{{\rm r}} = -0.36(1)$ and  $\eta_{{\rm b}} =- 0.70(1)$ that are report as solid lines in Fig.\,\ref{fig4}(c). Our experimental results interpolate between the single-atom case and the simulated 2D situation.

\section{Conclusion}\label{secV}

In summary we have studied the transmission of a macroscopic dense slab of atoms with uniform in-plane density and a transverse gaussian density distribution. We observed a strong reduction of the maximum optical density and a broadening of the resonance line. More surprisingly, we showed the presence of a large blue shift of the resonance line and a deviation from Lorentzian behavior in the wings of the resonance line. These results are qualitatively confirmed by coupled dipole simulations and a perturbative approach of this scattering problem. We also confirm the difficulty already observed to obtain a quantitative agreement between coupled dipole simulations and experimental results in the dense regime \cite{pellegrino2014observation,jennewein2016}. Possible explanations for this discrepancy are (i) residual motion of the atoms during the probing due to the strong light-induced dipole-dipole interactions, (ii) a too large intensity used in the experiment which goes beyond the validity of the coupled dipole approach, (iii) the influence of the complex atomic level structure. We were careful in this work to limit the influence of the two first explanations and the last possibility is likely to be the main limitation. The complex level structure leads to optical pumping effects during the probing and thus the scattering cross-section of the sample is not well-defined. A simple way to take into account the level structure is, as discussed in Sec.\,\ref{secIV}, to renormalize the scattering cross section by the average of the Clebsch-Gordan coefficients involved in the process. For $^{87}$Rb atoms this amounts for the factor 7/15 already discussed earlier. However this is a crude approximation which neglects optical pumping effects during scattering and whose validity in the dense regime is not clear. Two approaches can be considered to remove this limitation. First, one can use another atomic species such as strontium or ytterbium bosonic isotopes which have a spin singlet ground state and in which almost exact two-level systems are available for some optical transitions. Scattering experiments on strontium clouds have been reported \cite{Bidel02,kwong2014cooperative,Bromley16} but they did not explore the dense regime tackled in this work. The comparison with theory thus relies on modeling their inhomogeneous density distribution accurately. Second, an effective two-level system can be created in the widely used alkali atoms by imposing a strong magnetic field which could separate the different transitions by several times the natural linewidth as demonstrated in some recent experiments on three-level systems \cite{Whiting15,Whiting16}. This method could be in principle applied on our setup to create an effective two-level system and could help to understand the aforementioned discrepancies.

Finally, we note that this article focuses on the steady-state transmission of a cloud illuminated by a uniform monochromatic beam. The slab geometry that we have developed here is of great interest for comparison between theory and experiments and our work opens interesting perspectives for extending this study to time-resolved experiments, to fluorescence measurements or to spatially resolved propagation of light studies.

\vspace{0.2cm}
\begin{acknowledgments}
We thank Vitaly Kresin, Klaus M{\o}lmer, Janne Ruostekoski, Markus Greiner, Zoran Hadzibabic, Wilhelm Zwerger for fruitful discussions. This work is supported by DIM NanoK, ERC (Synergy UQUAM). L.C. acknowledges the support from DGA. This project has received funding from the European Union's Horizon 2020 research and innovation programme under the Marie Sk\l{}odowska-Curie grant agreement N$^\circ$ 703926.
\end{acknowledgments}

\appendix

\begin{figure}[t!!!]
\includegraphics[width=7.6cm]{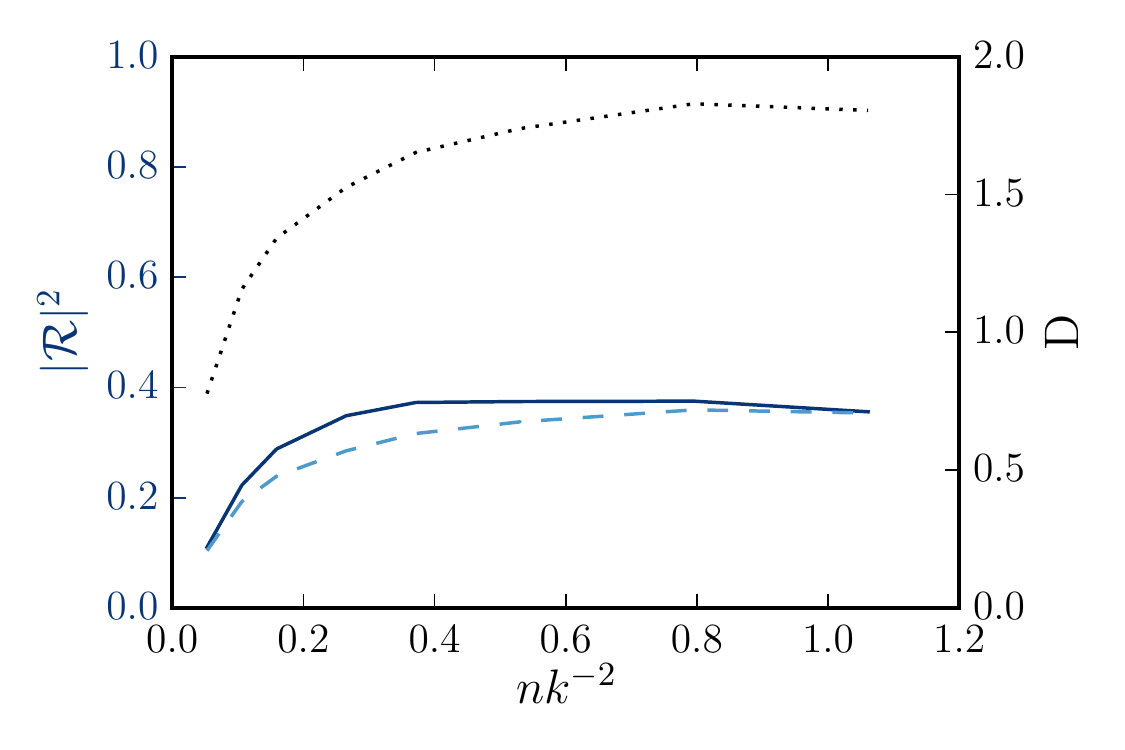}
\caption{Intensity reflection coefficient as a function of surface density for $k\Delta z=0$ (solid line). For comparison we show the corresponding optical depth $\mathrm D$ (dotted line, right axis) and the lower bound for the reflection coefficient deduced from this optical depth (dashed line). \label{fig8}}
\end{figure}

\section{Reflection coefficient of a 2D gas}

Thanks to their large scattering cross section at resonance, array of atoms can be used to emit light  with a controlled spatial pattern \cite{Jenkins12}. A single-atom mirror has been demonstrated \cite{Hetet11} and, more generally, regular two-dimensional arrays of atoms have been considered for realizing controllable light absorbers \cite{bettles2016enhanced} or mirrors  \cite{shahmoon2017cooperative} with atomic-sized thicknesses. For the disordered atomic samples considered in this article the strong decrease of the transmission because of dipole-dipole interactions could lead to a large reflection coefficient. For a strictly two-dimensional gas we show as a solid line in Fig.\,\ref{fig8} the result of the coupled dipole model for the  intensity reflection coefficient $|\mathcal{R}|^2$ at resonance and at normal incidence as a function of density. This intensity reflection coefficient has a behavior with density similar to the optical depth $\mathrm D$ (dotted line). The relation between these two quantities depends on the relative phase between the incoming and the reflected field. For a transmitted field in phase with the incident field we find, using the boundary condition $\mathcal R+ \mathcal T=1$, a lower bound for this reflection coefficient, $|\mathcal{R}|^2 \geq (1-|\mathcal T|)^2$, shown as a dashed line in Fig.\,\ref{fig8}. The intensity reflection coefficient is close to this lower bound in the regime explored in this work. The maximum computed value for the reflection coefficient is close to 40\,\% which shows that a single disordered layer of individual atoms can significantly reflect an incoming light beam \footnote{We have investigated here only the behavior at normal incidence. A full characterization of such a atomic mirror is beyond the scope of this work}. Note that for a non-2D sample light can be diffused at any angle. For our experimental thickness and the relevant densities the reflection coefficient is in practice much lower than the above prediction.

\bibliography{Bibliography}

\end{document}